\documentstyle[aaspptwo]{article}

\received{June 30, 1997}

\slugcomment{Submitted to The Astrophysical Journal Letters, 1997}

\begin{document}

\def\fd{IRAS F10214}
\def\fz{IRAS 09104}
\def\fq{IRAS F15307}
\def\mmu{$m^{-1}$}
\def\mmum{$m^{-1/2}$}
\def\hmd{$h^{-2}$}
\def\hmu{$h^{-1}$}

\title{ OLD MASSIVE ELLIPTICALS AND S0 IN THE HUBBLE DEEP FIELD
VANISH FROM VIEW AT $z>1.3$: POSSIBLE SOLUTIONS OF THE ENIGMA}


\author{Alberto Franceschini\altaffilmark{1}}
\affil{Dipartimento di Astronomia di Padova, Padova, Italy}

\author{Laura Silva\altaffilmark{2}}
\affil{International School for Advanced Studies}

\author{Gian Luigi\ Granato\altaffilmark{3}, 
Alessandro Bressan\altaffilmark{3}}
\affil{Osservatorio Astronomico di Padova, Padova, Italy}

\author{Luigi\ Danese\altaffilmark{2}}
\affil{International School for Advanced Studies}

\altaffiltext{1}{Dipartimento di Astronomia, Vicolo Osservatorio 5, I-35122,
Padova, Italy. E-mail: Franceschini@astrpd.pd.astro.it}
\altaffiltext{2}{International School for Advanced Studies, Via Beirut, 
I34014 Trieste, Italy}
\altaffiltext{3}{Osservatorio Astronomico, Vicolo dell'Osservatorio 5,
I35122 Padova, Italy. }

\begin{abstract}
We have investigated the properties of a bright K-band selected sample
of early-type galaxies in the Hubble Deep Field, as representative
of the field galaxy population. This dataset, based on public archives from
HST and from deep observations at Hawaii and Kitt-Peak, is unique as
for the morphological information on faint high-z sources, and for complete
photometric and spectroscopic coverage. The majority of bright
early-type galaxies in this field are found at redshifts $z \leq 1.3$
to share common properties with those of high-z cluster samples, 
as for the age and mass of the dominant stellar population --
which are found to be as old as 3-5 Gyr and as massive as $10^{11}\ M_\odot$
already at $z\simeq 1$. The bright end of the E/S0 population is already
in place by that cosmic epoch, with space densities, masses and
luminosities consistent with those of the local field population.
There is no evidence of a steepening of the mass function 
from $z$=0 to $z$=1, as inferred by some authors from analyses of 
optically-selected 
samples and favoured by hierarchical clustering models 
forming most of the E/S0s at $z<1$.
What distinguishes this sample is a remarkable absence of objects at
$z>1.3$, which would be expected as clearly detectable above the flux
limits, given the aged properties of the lower redshift counterparts.
So, something hide them at high redshifts. Merging 
could be an explanation, but it would require that already
formed and old stellar systems would assemble on short timescales of
1 Gyr or less. Our analysis shows that, during such a major assembling,
only a negligible fraction of the galactic mass in young stars has to be
added, a rather contrived situation from the dynamical point of view.
An alternative interpretation
could be that a dust-polluted ISM obscures the first 3 to 4 Gyr of the
evolution of these massive systems, after which a galactic wind 
and gas consumption makes them transparent. 
The presence of dust would have relevant (and testable)
implications for the evaluation of the global energetics from galaxy
formation and for the visibility of the early phases by current and
future infrared facilities. 

\end{abstract}


\section{INTRODUCTION}

Early-type galaxies have been studied in great detail in rich clusters
up to redshift one and above (see Dickinson, 1997), with the basic
result that passive evolution in luminosity and no evolution in mass is
ruling them at least up to $z=1.2$. Old ages and an early epoch of
formation are then implied by these observations.  It is however still
unclear how much representative of the general population is the
rich-cluster environment: according to a common wisdom, early-type
galaxies in the field might be much younger and follow a quite different
evolutionary pattern (Baugh et al. 1996; Kaufmann et al. 1996).

An exceedingly deep and clean view of field galaxy populations to high
redshifts is provided by a long integration of HST in the so-called {\it
Hubble Deep Field} (HDF, Williams et al., 1996).  Though having provided
important constraints on galaxy formation and evolution (Madau et al.
1997), the interpretation of these data is made difficult by the
relatively short selection wavelengths (essentially U, B, V, I), which
imply strong evolutionary and K-corrections as a function of redshift
and by the  sensitivity to even small amounts of
dust in the line-of-sight. The former two may be particularly severe for
the early-type galaxies, because of their quickly evolving optical
spectra and extreme K-corrections (e.g. Maoz, 1997).

Major arguments for performing time-expensive surveys in the
near-infrared, in particular in the K band, were to overcome most of
these problems of optical selection. A further one is that at these
wavelengths the fluxes from all (low-mass) stars dominating the baryonic
content of a galaxy are counted.

Various deep integrations have been performed in the HDF at near-IR
(Cowie et al. 1996; Dickinson, 1997) and mid-IR wavelengths
(Rowan-Robinson et al. 1997; Aussel et al. 1997), which, combined with
the extreme quality of the morphological information and the very good
spectroscopic coverage, make this area unique, in particular for the
investigation of early-type galaxies outside rich clusters.  

This letter is a preliminary report of an analysis of a bright complete
K-band selected sub-sample of elliptical and S0 galaxies in the HDF.
Section 2 discusses physical properties of the galaxies as inferred from
the morphology and from fits to the broad-band UV-optical-IR spectra.
Section 3 investigates the statistical properties of the sample (counts,
redshift distributions, identification statistics), and compare them
with those of local early-type galaxies. The distribution in space-time
of the sample is found to display a discontinuity at $z\sim1.3$, which
apparently conflicts with the relaxed and passively behaving nature of
the galaxies. Solutions of this paradox are discussed in Sections 4 and
5, by comparing the effects of merging with those of a long lasting
dust-extinguished phase.  We adopt $H_0=50\ Km/s/Mpc$ throughout the
paper.

\section{THE FIELD POPULATION OF EARLY-TYPE GALAXIES}

\subsection{The sample}

Our reference sample includes galaxies classified as ellipticals and S0
in the Hawaii Active Catalogue of the HDF by Cowie et al. (1997), having
K-band magnitudes brighter than 20 over the 4 square arcmin area covered
by the 3 HST-WF cameras, observed in the J and H+K by Cowie et al., 
and morphologically classified by van den Berg et al. (1996).  We have
re-defined the HK-selected sample reported by Songaila et al. into a
K-selected one by using the IRIM-KPNO image (Dickinson, 1997).
Magnitudes are measured in 5 arcsec diameter apertures.

The complete sample consists of 19 early-type galaxies with $K<20$, 
for 11 of which Cohen et al. (1996) report spectroscopic redshifts, while
for the remaining 8 a reliable estimate of $z$ is obtained from fits of
the optical-IR broad-band spectra. We have verified that the typical
error to be associated to the latter is not larger than 0.1.
Morphologically, these galaxies appear as quite regular early-type
systems up to the redshift limit of the sample, in no way dissimilar
from local counterparts. 

\epsfxsize=11cm
\begin{figure*}[ht]
\vspace*{-10pt}
\hspace*{0pt}
\epsffile{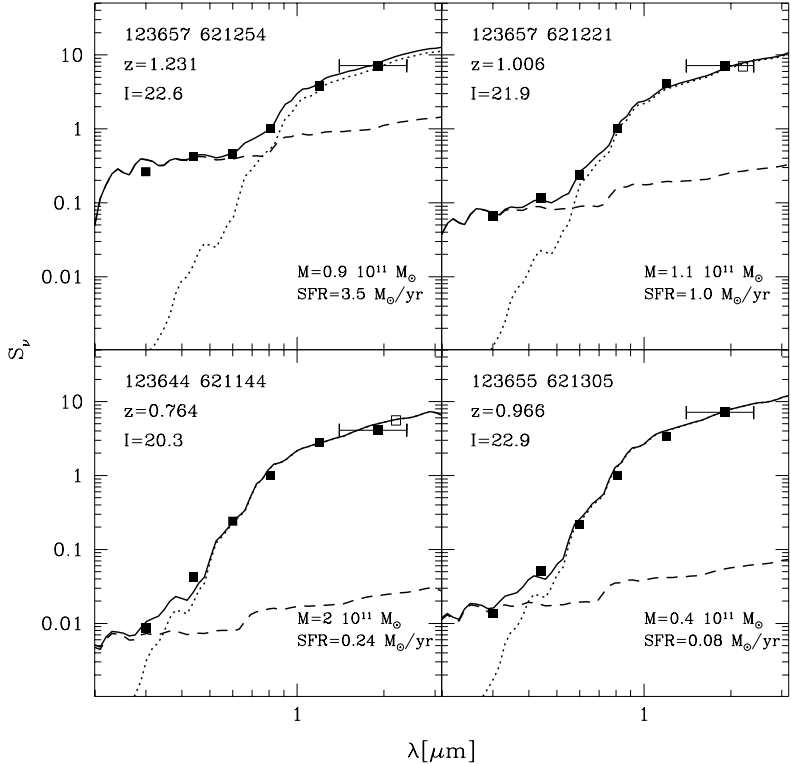}
\vspace*{-10pt}
\caption {Spectral energy distributions of four high-z early-type 
galaxies in the K-band HDF sample. Data on the source, the baryonic mass
and the star-formation rate are indicated.
}
\label{fig1}
\end{figure*}

The complete E/S0 sample to $K=20$ is supplemented by a fainter one
including 9 early-types and 7 unclassified objects down to $K=21$.
However, this enlarged sample has far less complete morphological
and spectroscopic information, and is then used only to test the models 
for consistency at fainter magnitudes and somewhat larger redshifts.

\subsection{Fitting the broad-band spectra}

The UV-optical-near/IR Spectral Energy Distributions (SED) for all
galaxies in the sample have been fitted with synthetic spectra based on
a model briefly summarized here.
%
The code follows two paths, one describing the chemical evolution 
of the galaxy's ISM, the other
associating to any galactic time a generation of stars with the
appropriate metallicity, and adding up the contributions 
light of all previous stellar populations.

The chemical path adopts a Salpeter IMF with a lower limit $M_l= 0.15\ 
M_\odot$, and a Schmidt-type law for the
Star Formation Rate (SFR): $\Psi(t)=\nu \, M_{g}(t)^{k}$, where $\nu$ is the
SF efficiency. A further parameter is the time-scale $t_{infall}$ for the 
infall of the primordial gas. The library of isochrones is 
based on the Padova models, spanning a
range in metallicities from $Z=0.008$ to 0.05 (i.e. from 0.3 to 2.5 solar). 
The isochrones are modified to account for the contribution by
dust-embedded Mira and OH stars during the AGB phase (Bressan et al.
1997), when circum-stellar dust reddens the optical emission and
produces an IR bump at 5-20 $\mu m$. This IR emission from AGB stars is
important for galactic ages of 0.1 to a few Gyrs. 

A number of evolutionary patterns for the time-dependent SFR $\Psi(t)$
have been tried to reproduce the galaxy SEDs (four examples of which
appear in Figure 1). Though solutions are not univocally determined, 
a few general trends are clearly written in the data. 
The first is that the bulk of the baryonic
mass in objects observed at the typical redshift $z\simeq 1$ has to be
older than at least three Gyr, under the assumption of an high average
stellar metallicity ($Z\simeq 2 Z_\odot$).  Lower metallicities would
imply bluer spectra, hence even older ages.

\epsfxsize=8.0cm
\begin{figure}[ht]
\vspace*{-10pt}
\hspace*{0pt}
\epsffile{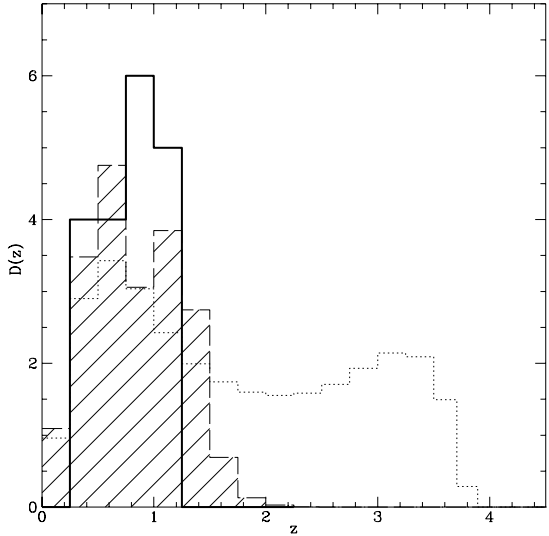}
\vspace*{-10pt}
\caption{The observed redshift distribution for sources identified as 
E/S0 in the K-HDF sample (thick line) versus predictions of model 1 (dotted 
line) and model 2 (shaded histogram). 
Most of the E/S0 galaxies are found at $z\geq 0.5$, because of K-correction 
and moderate evolution.
}
\label{fig2}
\end{figure}

The second evidence is that, overimposed on this old stellar component
dominating the spectra from 0.6 to 2.2 $\mu m$ and including typically
more than 90\% of the mass, a younger population is always present, as
suggested by the often blue U and B colours.  The evolution pattern
(model 1 hereafter) fitting the SEDs predicts that a large fraction of
the mass in stars has been generated during an intense burst lasting 0.7
Gyr, after which the input of energy by stellar winds and supernovae
produced a sudden outflow of the ISM through a galactic wind, stopping
the bulk of the SF. The evolution at later epochs is mostly due to
passive aging of already formed stellar populations.  
The dotted lines in Fig. 1, fitting the red side of spectra, are the 
contribution of such an old population (though not exactly
corresponding to model 1, but to a similar one described in Sect. 4
below). The global fit requires the additional
contribution of a minor fraction of the mass in younger stars 
(dashed lines) .

\section{STATISTICAL ANALYSIS OF A K-SELECTED SAMPLE: THE PROBLEM OF
THE REDSHIFT DISTRIBUTION}

We have shown that a dominant fraction of the galactic
baryonic mass has to be at least as old as 3 Gyr for objects observed at
$z\sim 1$.  Correspondingly, the luminosity of these stellar populations
is bound to increase with redshift
during most of the cosmic time preceeding the epoch when galaxies are
observed (the precise amount of such an increase depends
weakly on the detailed values of the model parameters).

It is of interest to check if this expected trend for the luminosity
evolution can be reconciled with the statistical properties 
of this and other K-selected galaxy samples.
To this end, model 1 has been combined with the the K-band
luminosity function (LLF) of galaxies to predict sample statistics,
under the assumption of a roughly constant galaxy mass function 
from redshift 0 to 3. Our adopted LLF is taken from
Gardner et al. (1997), which significantly updates previous
determinations. The separate contributions of early- (E/S0) and late-type 
(Sp/Ir) galaxies have been defined using the
morphological information on the optical LLF by Franceschini et al.
(1988) and type-dependent B-K colours.

We agree with Gardner et al. that the shape of the LLF can be naturally
reconciled with faint K-band counts only in an open universe, while a
closure one would require some "ad hoc" evolutionary prescriptions. Then
model 1 for E/S0 galaxies (see Sect. 2.2), 
supplemented with a moderate evolution for Sp/Ir galaxies as in Mazzei
et al. (1992), provides a good fit to the counts for $q_0=0.25$.

A crucial information is provided by the observed $z$-distribution
D(z) of early-type galaxies in the K-HDF  (Figure 2).  From $z=0$ to $z=1.2$, 
D(z) is consistent with a moderate (essentially passive) 
evolution in an open universe (with $q_0$=0.1-0.2). Again, a closure world 
model would require positive evolution, i.e. a larger comoving 
luminosity/density of massive E/S0s at $z$=1 than locally.

Hence, there is no evidence, up to this epoch, of an evolution of the baryon
mass function. This is difficult to reconcile with a decrease of the mass 
and luminosity due
to progressive disappearence of big galaxies in favour of smaller mass 
units, as implied by some specifications of the hierarchical clustering 
scheme (e.g. Baugh, Cole, \& Frenk, 1996). This also seem at odd with the
finding by Kauffmann, Charlot \& White (1996) of a strong negative 
evolution of the early-type population
(by a factor 2-3 less in number density at $z$=1). 

Above z=1.2, however, model 1 has difficulties to reproduce the
redshift distribution reported in Fig. 2 and the identification 
statistics in Figure 3. 
A remarkable feature of the observational D(z) is the sharp cutoff
at $z>1.3$, such that no galaxies of the complete sample occur at such
$z$. Model 1 is in no way able to explain the large number of sources  
at $z=1$ followed by the rapid convergence above: the
prediction would be of a much more gentle distribution, with less a
pronounced peak and a large tail of galaxies (incuding typically half of
sample objects) observable above $z=1.2$. 
Lower values of $q_0$ would allow better fits of the observed D(z) up 
to $z=1.3$, but would also worsen the mis-fit at higher $z$.

A similar effect is observed in the
counts as a function of the morphological class (Fig. 3). Here again
sources identified as E/S0s show steep counts to K=19 and a sudden
convergence thereafter, while the model would predict much less rapid
change in the slope. These problems remain forcing in various ways the 
model parameters.

\epsfxsize=8.0cm
\begin{figure}[ht]
\vspace*{-10pt}
\hspace*{0pt}
\epsffile{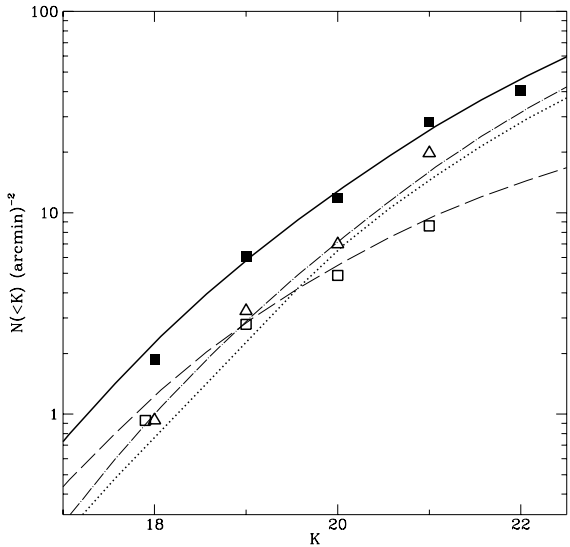}
\vspace*{-10pt}
\caption{Identification statistics for E/S0 (open squares)
and Sp/Ir (open triangles) in the K-HDF, as a function of magnitude limit.
Filled squares are the total observed counts. The dot-dashed line is the
prediction for Sp/Ir, the dotted line is for E/S0s after model 1, while the
long-dashed line for E/S0s after model 2. The thick continuous line is the
prediction of model 2 for the total counts.
}
\label{fig3}
\end{figure}

\section{EXPLAINING THE ENIGMA: DUST EXTINTION DURING A PROLONGED
ACTIVITY PHASE}

\subsection{Effects of merging at $z>1.3$}

At redshifts larger than 1.3, early-type galaxies suddenly disappear 
from the K-selected sample, while we would expect to observe them to much
higher $z$, given the predicted luminosity enhancement when approaching
the most intense phase of star formation. The latter is expected to happen 
at $z$=2 to 5 after our analysis of the stellar populations in $z$=1 objects. 

In principle, this lack of high-z objects might be taken as an
indication in favour of a fast {\it merging}-driven evolution of the mass
function of galaxies at these redshifts.
To be effective, however, this process has to operate on
short timescales ($\leq 1$ Gyr) and, in particular, to assemble
already formed, old and metal rich stellar populations, as implied by
the spectra of $z\sim 1-1.2$ galaxies (see Fig. 1). 
Only a small fraction (less than a few percent) of the mass in young stars 
has to be added to the galaxy, so stellar systems
essentially deprived of gas are to be accreted, otherwise the gas would make 
stars actively during this dynamically perturbed phase.  
As suggested by N-body   
simulations, typical outcomes of strong galaxy collisions
are rings and bars, which are very rarely observed in HDF galaxies (van
den Berg et al., 1996).

Finally, it seems unlikely that the {\it merge} producing ellipticals
and S0s should operate only among gas-poor systems, if we consider that
gas-rich objects are so numerous in the field.

\subsection{Effects of dust during a prolonged SF phase in field
ellipticals}

An alternative interpretation, discussed here, for the lack of objects
at $z>1.3$ in the K-HDF assumes that dust, present in a highly
metal-enriched ISM during the first 3 to 4 Gyr of the evolution of
massive galaxies, extinguishes the light emitted by high redshift objects.
The idea is that the SF efficiency in field galaxies with deep dark-matter 
potential wells is not so high to produce a galactic wind on
short timescales after the onset of SF. Hence an ISM is present in the 
galaxy for an appreciable fraction of the Hubble time, during which the ISM 
is progressively enriched of metals and dust. While the gas fraction
diminishes as stars are continuously formed at low rates, its
metallicity increases and keeps the dust optical depht roughly constant
with time. 

We have then extended the model of Sect. 2.2 to account for
the effect of extinction and re-radiation by dust in star-forming
molecular regions and more diffused in the galaxy body (see for details 
Granato et al. 1997).
The amount of dust in the galaxy is assumed proportional to the residual gas
fraction and chemical abundance of C and Si, while the
relative amount of molecular to diffuse gas ($M_m/M_g$) is a model parameter.
The molecular gas is divided into spherical clouds of assigned mass
($5\ 10^{5}\; \rm{M}_{\odot}$) and radius ($16\ pc$).  
Each generation of stars born within the cloud progressively
escapes it on timescales of a few Myr with a velocity $v_{out}$.
The emerging spectrum is obtained by
solving the radiative transfer through the cloud.
Before escaping the galaxy, the light arising from young stars/molecular
complexes, as well as from older stellar generations, further interacts
with dust present in the diffuse gas component (the latter dominates the 
global galactic extinction at late epochs, when the SF rate is low).
The dimming of starlight and consequent dust emission are
computed by describing the galaxy as a spherically symmetric system
subdivided into volume elements, with radial dependences of stars and gas 
described by a King profile with a core radius $r_c=400\ pc$.

This evolutionary code was used to explore alternatives to model 1.
We found a  solution (hereafter model 2) with the following parameter values:
baryonic mass $M=10^{11}M_\odot$, (typical for our sample
galaxies), $t_{infall}=0.1$ Gyr, $k=1.5$, $\nu=1.3$, $M_m/M_g=0.3$, 
$v_{out}=0.2\ Km/s$. Correspondingly, the SFR $\Psi(t)$
has a peak value of $\sim 100\ M_\odot/yr$ at galactic time $t=0.2$ Gyr
and then decays as a power-law $\propto t(Gyr)^{-1.7}$. After 4.5 Gyr,
the balance of gas thermal energy vs. input from supernavae breaks, and
formally most of the small residual gas is then lost by the galaxy. 
After this event a low-level SF activity keeps on, due to either
partial efficiency of the wind, or to stellar re-cycling.
The average stellar metallicity of the remnant is $Z=2.5 Z_\odot$.

This scheme naturally accounts for the evidence (Fig. 1) that the
highest redshift galaxies in the K-HDF show a combination of a massive
amount of old ($>3$ Gyr) stellar populations plus a small fraction 
of younger stars.
Fits of galaxy SEDs from model 2 (as reported in Fig. 1) can be found
assuming an open world model with $q_0=0.2$. The redshifts $z_F$ for the
onset of SF span a fairly wide range from 2 to 5 (field early-types are 
not coeval in this sense, and in particular low-mass systems are younger
on average), with an average value $z_F=3.5$. 

\epsfxsize=8.5cm
\begin{figure}[!ht]
\vspace*{-10pt}
\hspace*{0pt}
\epsffile{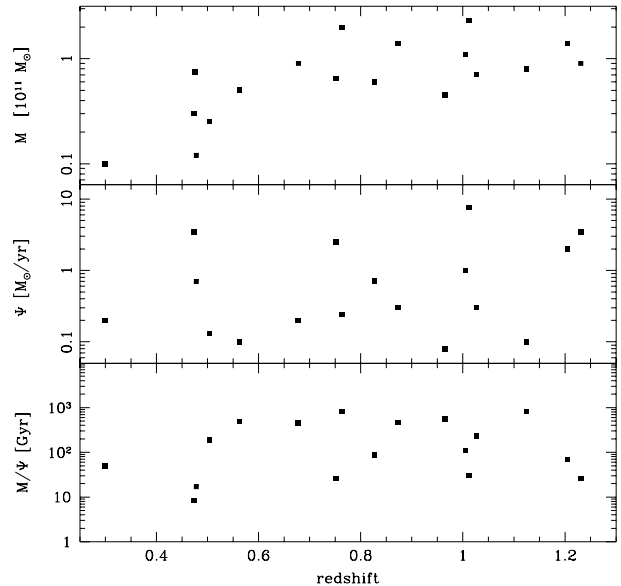}
\vspace*{-10pt}
\caption{Basic properties of the E/S0 sample as inferred from best-fitting
model 2 on the observed SEDs. Top panel is an estimate of the total
baryonic mass; mid panel is the ongoing star formation rate; bottom panel
is the ratio of the two, providing an estimate of the time needed
to synthesize that mass in stars with that rate of SF.
}
\label{fig4}
\end{figure}

By constraction, the same scheme is successful in reproducing statistical 
distributions in Figs. 2 and 3, in addition to other galaxy counts in the
K-band. Now both the drastic cutoffs of D(z) for E/S0s at z=1.3 and of
the counts at $K>19$ are reproduced.

Figure 4 summarizes our understanding of the E/S0 sample galaxies, as
for the baryonic mass (top), the residual SF rate $\Psi$
(mid), and the timescale needed by this SF activity to build the
observed mass in stars (bottom). The residual SF rate is typically $<1\
M_\odot/yr$, and  
cannot contribute more than a few \% of the baryonic mass for most of
the galaxies. Note, finally, that the results in Fig. 4 are very robust:
essentially the same would be found using model 1 instead of 2.

\section{CONCLUSIONS AND PERSPECTIVES}

Near-IR observations provide a view of galaxy evolution minimally biased
by the effects of the evolutionary K-correction, dust extinction and
changes in the $M/L$ ratio due to the aging of stellar populations. We
have used a near-IR sample in the HDF with optimal morphological
information and spectro-photometric coverage to study 
properties of distant early-type galaxies outside rich clusters. 

The SEDs of early-types in the HDF can
be explained in terms of a dominant old stellar population (older than
3-5 Gyr already at z=1) with a mass of $10^{11}\ M_\odot$. The residual
SF indicated by the blue U-B colours can only contribute some percent
of the mass in stars. So, the main SF event producing the bulk of stars
is confined to happen at $z>2$ for reasonable world models.
These general properties are not inconsistent with those found for
cluster ellipticals studied at medium to high redshifts.
Early-types in the field are not coeval, however, 
the formation redshifts spanning a fairly wide range, $z_F\simeq 2$ to 5,
and the low-mass systems being younger on average. 

A basic problem is found when comparing the fraction of high-z objects
expected on the basis of a simple photometric model
with the observed distribution: a sizeable number would be expected
at $z>1.2$, while none is observed.
Though not all redshifts in the sample are spectroscopically confirmed,
the latter conclusion seems quite robust:
if optical spectroscopy has difficulties to enter this redshift
domain, our analysis, using broad-band galaxy SEDs when spectroscopic
redshifts are not available, has no obvious biases,
while the uncertainty in the photometric estimate of $z$ for this kind
of spectra is small.
Of course, caution is in order about our conclusions because of the limited
statistics. Improvements will be achieved by further areas deeply
investigated by HSF (e.g. the southern HDF) and by further
efforts of optical and near-IR spectroscopy.

\epsfxsize=8.5cm
\begin{figure}[!ht]
\vspace*{-10pt}
\hspace*{0pt}
\epsffile{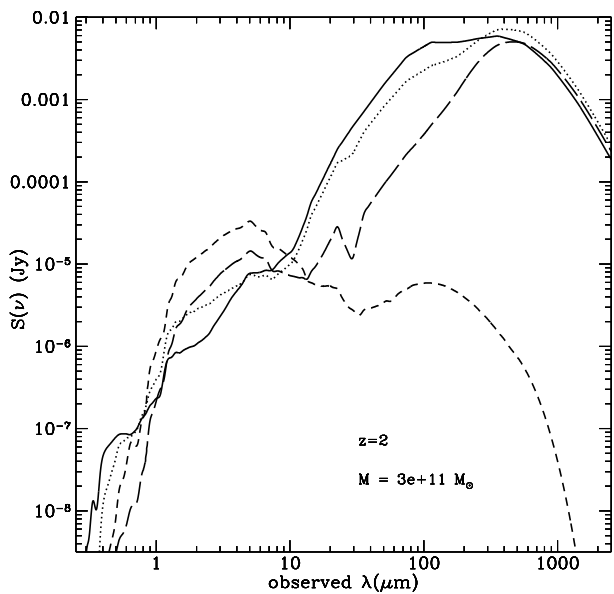}
\vspace*{-10pt}
\caption{Predicted spectra of a $M=3\ 10^{11}\ M_\odot$ galaxy at $z=2$
observed at 0.5 (continuous), 1 (dotted), 4 (long dash), and 5 (short 
dash) Gyr, according to model 2.
}
\label{fig5}
\end{figure}

To explain the large number of E/S0s at z=1 and their quick disappearence
above,
{\it merging} would have to operate on quite short timescales ($\leq 1$ 
Gyr) and to assemble mostly stellar systems, to avoid triggering an 
excess amount of star formation. This is perhaps an uneasy situation as for
the dynamics and for the probability of the event, that would be
expected to favour, if any, dynamically efficient encounters between the
numerous gas-rich systems. 

We have discussed an alternative solution in terms of a somewhat protracted
presence of a dusty and opaque ISM during the first 3-4 Gyr of galaxy's
life, after which thermal unbalance in the ISM - due to a low residual 
gas mass and energy input by type-I SN - would clean up the object
and make it transparent. A somewhat protracted SF activity in field, as 
opposed to cluster, ellipticals may be consistent with their observed
"disky" morphology and the $H_\beta$ narrow-band indices.

The implication of this is that long
wavelength observations would be better suited to detect and
characterize the early evolutionary phases of spheroidal galaxies.
Figure 5 illustrates possible spectra corresponding to different ages of
a $M=3\ 10^{11}\ M_\odot$ galaxy at $z=2$, according to model 2.
As indicated by the figure, various planned missions
(in launch-time order, SIRTF, ESA's FIRST,
NASA's NGST, and some ground-based observatories in
exceptionally dry sites, like the South Pole) could discover unexpectedly
intense activity of star-formation at $z=1.5$ to 5.
Also interesting tests of these ideas could be soon achieved with SCUBA 
on JCMT, and with ISO.

%
%

\acknowledgments
This work has been funded by EC TMR Network Programmes FMRX-CT96-0068
and MRX-CT96-0086.



\begin{thebibliography}{}

\bibitem{}{}
Aussel, H., Cesarsky, C., Elbaz, D., Stark, J.L. 1997, to appear in 
"Extragalactic IR Astronomy", Moriond 1997

\bibitem{}{}
Baugh, C., Cole, S., Frenk, C. 1996, MNRAS, 283, 1361

\bibitem{}{}
Bressan, A., Granato, G.L., Silva, L. 1997, AA submitted

\bibitem{}{}
Cohen, J., Cowie, L., Hogg, D., Songaila, A., Blandford, R., Hu, E., 
Shopbell, P. 1996, AJ, 471, L5

\bibitem{}{}
Cowie, L., et al. 1996, ~~~~~~~~~~~~~~~~~~~~~
\verb+ http://www.ifa.hawaii.edu/~cowie/k_table.html+

\bibitem{}{}
Dickinson, M. 1997, Preprint Astro-ph 9703035

\bibitem{}{}
Dickinson, M. 1997, ~~~~~~~~~~~~~~~~~~~~~~~~~~~~~~~~
\verb+ http://archive.stsci.edu/hdf/irim.html +

\bibitem{}{}
Franceschini A., Danese L., De~Zotti G., Toffolatti L. 1988, MNRAS, 233, 
157 

\bibitem{}{}
Gardner, J., Sharples, R., Frenk, C., Carrasco, B. 1997, NASA Preprint 
97-4

\bibitem{}{}
Giavalisco, M., Steidel, C., Macchetto

\bibitem{}{}
Granato, G.L., et al. 1997, in preparation

\bibitem{}{}
Kaufmann, G., Charlot, S., White, S. 1996, MNRAS, 283, L117

\bibitem{}{}
Maoz, D. 1997, Preprint Astro-ph 704173

\bibitem{}{}
Rowan-Robinson, M. 1997, MNRAS, in press

\bibitem{}{}
Williams, R., et al. 1996 AJ, 112, 1335

\bibitem[]{} 
Madau P., Ferguson H.C., Dickinson M.E., Giavalisco M., Steidel C.C.,  
Fruchter A., 1996, MNRAS, 283, 1388

\bibitem{}{}
Mazzei, P., Xu, C., De Zotti, G. 1992, AA, 256, 45

\bibitem{}{}
van den Berg, S., Abraham, R., Ellis, R., Tanvir, N., Santiago, B.,
Glazebrook, K. 1996, AJ, 112, 359

\end{thebibliography}

\end{document}